\documentclass[preprint,showpacs,preprintnumbers,amsmath,amssymb]{revtex4}
\usepackage{amsmath,amsfonts,amssymb}
\usepackage{epsfig}
\def\bit{\begin{itemize}}
\def\eit{\end{itemize}}
\def\ben{\begin{enumerate}}
\def\een{\end{enumerate}}
\def\bed{\begin{description}}
\def\eed{\end{description}}
\def\b{\beta}

\def\k{\kappa}
\def\l{\lambda}

\def\q{\quad}

\def\half{\frac{1}{2}\,}
\def\third{\frac{1}{3}\,}
\def\quart{\frac{1}{4}\,}
\def\lsim{\raise0.3ex\hbox{$<$\kern-0.75em\raise-1.1ex\hbox{$\sim$}}}
\def\gsim{\raise0.3ex\hbox{$>$\kern-0.75em\raise-1.1ex\hbox{$\sim$}}}

\let\jnfont=\rm
\def\NPB#1,{{\jnfont Nucl.\ Phys.\ B }{\bf #1},}
\def\PLB#1,{{\jnfont Phys.\ Lett.\ B }{\bf #1},}
\def\EPJC#1,{{\jnfont Eur.\ Phys.\ Jour.\ C }{\bf #1},}
\def\PRD#1,{{\jnfont Phys.\ Rev.\ D }{\bf #1},}
\def\PRL#1,{{\jnfont Phys.\ Rev.\ Lett.\ }{\bf #1},}
\def\MPLA#1,{{\jnfont Mod.\ Phys.\ Lett.\ A }{\bf #1},}
\def\JPG#1,{{\jnfont J.\ Phys.\ G}{\bf #1},}
\def\CTP#1,{{\jnfont Commun.\ Theor.\ Phys.\ }{\bf #1},}
\def\JHEP#1,{{\jnfont JHEP \ }{\bf #1},}
\def\NPPS#1,{{\jnfont Nucl.\ Phys.\ Proc.\ Suppl.\ }{\bf #1},}

\def\beq{\begin{equation}}
\def\eeq{\end{equation}}
\def\bea{\begin{eqnarray}}
\def\eea{\end{eqnarray}}
\newcommand{\ba}{\begin{array}}
\newcommand{\ea}{\end{array}}
\def\nn{\nonumber}

\begin{document}
\title{Dark Matter in the Singlet Extension of MSSM:\\
        Explanation of Pamela and Implication on Higgs Phenomenology}

\author{Wenyu Wang$^1$, Zhaohua Xiong$^1$,
                  Jin Min Yang$^2$, Li-Xin Yu$^2$}

\affiliation{
$^1$ Institute of Theoretical Physics, College of Applied Science,
              Beijing University of Technology, Beijing 100020, China\\
$^2$  Key Laboratory of Frontiers in Theoretical Physics,
      Institute of Theoretical Physics, Academia Sinica,
              Beijing 100190, China   \vspace*{1.5cm}}

\begin{abstract}
As discussed recently by Hooper and Tait, the singlino-like dark
matter in the Minimal Supersymmetric Standard Model (MSSM)
extended by a singlet Higgs superfield can give a perfect
explanation for both  the relic density and  the Pamela result
through the Sommerfeld-enhanced annihilation into singlet Higgs
bosons ($a$ or $h$ followed by $h\to a a$) with $a$ being light
enough to decay dominantly to muons or electrons. In this work we
analyze the parameter space required by such a dark matter
explanation and also consider the constraints from the LEP
experiments. We find that although the light singlet Higgs bosons
have small mixings with the Higgs doublets in the allowed
parameter space, their couplings with the SM-like Higgs boson
$h_{SM}$ (the lightest doublet-dominant Higgs boson) can be
enhanced by the soft parameter $A_\kappa$ and, in order to meet
the stringent LEP constraints, the $h_{SM}$ tends to decay into
the singlet Higgs pairs $aa$ or $hh$ instead of $b\bar b$. So the
$h_{SM}$ produced at the LHC will give a multi-muon signal,
$h_{SM}\to aa\to 4 \mu$ or $h_{SM}\to hh\to 4 a \to 8 \mu$.
\end{abstract}

\maketitle
\section{Introduction}
The experiment Pamela has observed an excess of the cosmic ray
positron in the energy range 10-100 GeV ~\cite{pamela}, which is
hard to be explained by the conventional cosmic ray source
\cite{secbg}. While there may exist some mundane explanations like
pulsars~\cite{pulsars} and the acceleration of positron
secondaries in cosmic ray acceleration regions~\cite{secacc}, the
dark matter interpretation ~\cite{dm-int,sommerfeld2} is
especially interesting since it may be related to new physics to
be probed at the LHC.

To explain the Pamela excess by the dark matter annihilations, there
are some challenges.
First, the dark matter must annihilate dominantly into leptons since Pamela
has observed no excess of anti-protons ~\cite{pamela} (However, as pointed 
in \cite{kane}, this statement may be not so solid due to the significant 
astrophysical uncertainties associated with their propagation).
Second, the explanation of Pamela excess requires an annihilation rate which
is too large to explain the relic abundance if the dark matter
is produced thermally in the early universe.
To tackle these difficulties, a new theory of dark matter was proposed in \cite{sommerfeld2}.
In this new theory the Sommerfeld effect of a new force in the dark sector can
greatly enhance the annihilation rate when the velocity of dark matter is much smaller
than the velocity at freeze-out in the early universe, and the dark matter annihilates
into light particles which are kinematically allowed to decay to muons or electrons.

The above fancy idea is hard to realize in the popular Minimal Supersymmetric Standard Model
(MSSM) because there is not a new force in the neutralino dark matter sector to
induce the Sommerfeld enhancement and the neutralino drak matter annihilates largely to final
states consisting of heavy quarks or gauge and/or Higgs bosons~\cite{jungman,neu}.
However, as discussed in \cite{Hooper:2009gm}, in the extension of the MSSM by
introducing a singlet Higgs superfield, the idea in \cite{sommerfeld2} can be
realized by the singlino-like neutralino dark matter (hereafter the singlino-like neutralino
is simply called singlino):
\begin{itemize}
\item[(i)] The singlino dark matter annihilates to the light singlet Higgs bosons
and the relic density can be naturally obtained from the
interaction between singlino and singlet Higgs bosons;
\item[(ii)] The singlet Higgs bosons, not related to electroweak symmetry breaking,
can be light enough to be kinematically allowed to
decay dominantly into muons or electrons through the tiny mixings with the Higgs doublets;
\item[(iii)] The Sommerfeld enhancement needed in the dark matter annihilation for the
explanation of Pamela result can be induced by the light singlet Higgs boson ($h$).
\end{itemize}
Such an explanation of dark matter requires that
the singlet Higgs field has
very small mixing with the Higgs doublets,
which implies that the singlino dark matter may remain hidden and irrelevant
to the LHC experiments.
However,
we note that the singlet extension of the MSSM has a quite large parameter space
and thus the
coupling of the light singlet Higgs ($h$, $a$) with the doublet Higgs
(the lightest one is called $h_{SM}$) may be enhanced by other parameters.
For example, through the soft term
$A_\kappa S^3$ ($S$ is the singlet Higgs field) with a large $A_\kappa$,
a pair of singlet Higgs bosons may sizably couple to a doublet Higgs boson although
the mixing between the singlet and doublet Higgs fields is small.
Therefore, this model may allow for exotic Higgs phenomenology at the LHC.

In this work we study the parameter space allowed by the
explanation of Pamela result plus relic density via Sommerfeld
enhancement and also consider the constraints from the LEP
experiments. We find that although the light singlet Higgs bosons
have small mixings with the Higgs doublets, their couplings with
the SM-like Higgs boson ($h_{SM}$) can be enhanced by the soft
parameter $A_\kappa$ and, in order to meet the stringent LEP
constraints, the $h_{SM}$ tends to decay into the singlet Higgs
pairs $aa$ or $hh$ instead of $b\bar b$. This implies that the
$h_{SM}$ produced at the LHC will give a multi-muon signal,
$h_{SM}\to aa\to 4 \mu$ or $h_{SM}\to hh\to 4 a \to 8 \mu$.

This work is organized as follows. In Sec. \ref{sec2} we discuss
the Higgs and neutralino sectors in the singlet extension of the MSSM.
In Sec. \ref{sec3} we scan the parameter space allowed by the dark matter explanation
and LEP experiments, and discuss the implication on Higgs phenomenology.
Finally, a summery is given in Sec. \ref{sec4}.

\section{Higgs and neutralinos in singlet extention of MSSM}
\label{sec2}
The Higgs superpotential in the general singlet extension of the MSSM is given 
by \cite{Hooper:2009gm}
\bea
W = \mu \widehat{H}_u \cdot \widehat{H}_d+\l \widehat{S}
\widehat{H}_u \cdot \widehat{H}_d+\eta \widehat{S}
+\half \mu_s \widehat{S}^2 + \frac{1}{3} \k \widehat{S}^3 \ ,
\label{superpotential}
\eea
where $\widehat{S}$ is the singlet Higgs superfield while $\widehat{H}_u$
and  $\widehat{H}_d$ are the doublet Higgs superfields.
The Higgs scalar potential consists of the D-term, the F-term and the soft SUSY-breaking term.
Since $\widehat{S}$ is a singlet, the D-term is same as in
the MSSM.  The F-term from the superpotential is given by
\bea
V_F=|\mu+\l S|^2(|H_u|^2 + |H_d|^2)
                            +|\eta+\mu_s S +\l H_u\cdot H_d+\k S^2|^2.\label{vfud}
\eea
The soft SUSY-breaking terms are given by
\bea
V_{\rm soft}&=& m_\mathrm{H_u}^2 | H_u |^2 + m_\mathrm{H_d}^2 | H_d |^2
    + m_\mathrm{S}^2 | S |^2 \nn\\
&& +(B \mu H_u \cdot H_d +\l A_\l\ H_u \cdot H_d S
   + C\eta S +\half B_s \mu_s S^2 + \third \k A_\kappa\ S^3 + \mathrm{h.c.})\,.
\eea
So the Higgs potential reads
\bea
V &=& |\mu+\l S|^2(|H_u|^2 + |H_d|^2)+ |\l H_u\cdot H_d+\k S^2|^2\nn\\
&& +\quart g^2 (|H_u|^2 - |H_d|^2)^2 + \half g_2^2 |H_u^+H_d^{0*} +H_u^0 H_d^{-*}|^2\nn\\
&& +m_\mathrm{H_u}^2|H_u|^2 + m_\mathrm{H_d}^2|H_d|^2 + m_S^2|S|^2\nn \\
&& +(B\mu H_u\cdot H_d + \l A_\l H_u \cdot H_d S + \l\mu_s H_u\cdot H_d S^\ast\nn\\
&& +C\eta S+\half B_s \mu_sS^2+ \third \k A_\k\ S^3 +\k\mu_s S^2 S^\ast + \mathrm{h.c.})
\label{vform}
\eea
where $g^2 = (g_1^2 + g_2^2)/2$ with $g_1$ and $g_2$ being respectively the coupling constant
of SU(2) and U(1) in the SM.

After the Higgs fields develop the vevs $h_u$, $h_d$ and $s$, i.e.,
\beq
H_u^0 = h_u + \frac{H_{uR} + iH_{uI}}{\sqrt{2}} , \q
H_d^0 = h_d + \frac{H_{dR} + iH_{dI}}{\sqrt{2}} , \q
S = s + \frac{S_R + iS_I}{\sqrt{2}}
\eeq
we obtain a $3\times 3$ mass matrix ${\cal M}_h$ for CP-even Higgs bosons,
a $3\times3$ mass matrix ${\cal M}_a$ for CP-odd Higgs bosons and a
$2\times 2$ mass matrix ${\cal M}_c$ for the charged Higgs bosons:
\begin{itemize}
\item[(1)] The CP-even Higgs mass matrix in the basis $(H_{uR}, H_{dR}, S_R)$ is given by
\bea
{\cal M}_{h,11} & = & g^2 h_u^2 + \cot\beta\left[\l s (A_\l + \k s+\mu_s) +B\mu\right], \\
{\cal M}_{h,22} & = & g^2 h_d^2 + \tan\beta\left[\l s (A_\l + \k s+\mu_s) +B\mu\right], \\
{\cal M}_{h,33} & = & \l (A_\l+\mu_s) \frac{h_u h_d}{s}\, -\l \frac{\mu}{s} (h_u^2+h_d^2)
                      + \k s (A_\k + 4 \k s+3\mu_s)-\frac{C\eta}{s},\\
{\cal M}_{h,12} & = & (2\l^2 - g^2) h_u h_d - \l s (A_\l + \k s +\mu_s)-B\mu, \\
{\cal M}_{h,13} & = & 2\l (\mu+\l s) h_u  - \l h_d (A_\l + 2\k s+\mu_s), \\
{\cal M}_{h,23} & = & 2\l (\mu+\l s) h_d - \l h_u (A_\l + 2\k s+\mu_s),
\eea
where $\tan\beta = h_u/h_d$. This mass matrix can be diagonalized by a rotation
\bea
\left( \begin{array}{l} h_1 \\ h_2 \\ h_3 \end{array} \right) 
 = U \left(\begin{array}{l} H_{uR} \\ H_{dR} \\S_R \end{array} \right)
\label{rotation} 
\eea
with an orthogonal matrix $U$.
The mass eigenstates are ordered as $m_{h_1}<m_{h_2}<m_{h_3}$.
In the MSSM limit ($\l$, $\eta$, $\mu_s$, $\k \to 0$ and $h_3 \sim S_R$)
the elements of the first $2 \times 2$ sub-matrix of $U$ are
related to the MSSM angle $\alpha$ as
\bea
&&U_{11} =~~ \cos\alpha\ , \qquad U_{21} = \sin\alpha\ ,\nn \\
&&U_{12} =-\sin\alpha\ , \qquad U_{22} = \cos\alpha\ .
\eea
\item[(2)] The CP-odd Higgs mass matrix ${\cal M}_a$ in the basis $(H_{uI}, H_{dI}, S_I)$
is given by
\bea
{\cal M}_{a,11} & = & \cot\beta[\l s (A_\l + \k s+\mu_s)+B\mu], \\
{\cal M}_{a,22} & = & \tan\beta[\l s (A_\l + \k s+\mu_s)+B\mu], \\
{\cal M}_{a,33} & = & 4 \l \k h_u h_d + \l (A_\l+\mu_s) \frac{h_u h_d}{s} \nn\\
              && -\l\frac{\mu}{s}(h_u^2+h_d^2)
                 -\k s(3A_\k+\mu_s)-\frac{C\eta}{s}-2B_s\mu_s, \\
{\cal M}_{a,12} & = & \l s (A_\l + \k s+\mu_s)+B\mu, \\
{\cal M}_{a,13} & = & \l h_d (A_\l - 2\k s-\mu_s), \\
{\cal M}_{a,23} & = & \l h_u (A_\l - 2\k s-\mu_s).
\eea
The diagonalization of this mass matrix can be performed in two steps.
The first step is to rotates into a basis ($\tilde{A}, \tilde{G}, S_I$)
with $\tilde{G}$ being a massless Goldstone mode:
\beq
\left(\ba{c}H_{uI} \\  H_{dI} \\ S_I \ea\right) =
 \left(\ba{ccc} \cos\b & -\sin\b & 0 \\
 \sin\b & \cos\b & 0 \\
 0 & 0 & 1 \ea\right)
\left(\ba{c} \tilde{A} \\ \tilde{G} \\  S_I \ea\right).
\eeq
Dropping the Goldstone mode, the
remaining $2 \times 2$ mass matrix in the basis ($\tilde{A}, S_I$) is given by
\bea
{\cal M}_{a,11} & = & (\tan\beta+\cot\beta)[\l s (A_\l + \k s+\mu_s)+B\mu], \\
{\cal M}_{a,22} & = & 4 \l \k h_u h_d + \l (A_\l+\mu_s) \frac{h_u h_d}{s}
             -\l\frac{\mu}{s}(h_u^2+h_d^2) \nn \\
             && -\k s(3A_\k+\mu_s)-\frac{C\eta}{s}-2B_s\mu_s, \\
{\cal M}_{a,12} & = & \l \sqrt{h_u^2+h_d^2}\, (A_\l - 2\k s-\mu_s).
\eea
It can be diagonalized by an orthogonal $2 \times 2$ matrix
$P'$ and the physical CP-odd states $a_i$ are given by
(ordered as $m_{a_1}<m_{a_2}$)
\bea
a_1 &=& P_{11}' \tilde{A} + P_{12}' S_I
    = P_{11}' (\cos\b H_{uI} + \sin\b H_{dI} ) +P_{12}'S_I ,  \\
a_2 &=& P_{21}' \tilde{A} + P_{22}' S_I
   = P_{21}' (\cos\b H_{uI} + \sin\b H_{dI} ) +P_{22}'S_I ,
\eea
\item[(3)] The charged Higgs mass matrix ${\cal M}_\pm$ in the basis
$\left(H_u^+, H^+_d\right)$ is given by
\beq
{\cal M}_\pm = \left(\l s (A_\l + \k s+\mu_s) + B\mu + h_u h_d (\frac{g_2^2}{2} - \l^2)\right)
\left(\ba{cc} \cot\b & 1 \\ 1 & \tan\b \ea\right) ,
\eeq
which gives one eigenstate $H^\pm$ of mass Tr${\cal M}_\pm$ and
one massless goldstone mode $G^\pm$:
\bea
H_u^\pm &=& \cos\b H^\pm - \sin\b G^\pm\ ,\nn \\
H_d^\pm &=& \sin\b H^\pm + \cos\b G^\pm\ .
\eea
\item[(4)] The neutralino mass matrix ${\cal M}_0$ can be read from the Lagrangian
\bea
{\cal L} & = & \half M_1 \l_1 \l_1 + \half M_2 \l_2^3 \l_2^3 \nn \\
&& +\mu\psi_u^0 \psi_d^0 + \l (s \psi_u^0 \psi_d^0 + h_u \psi_d^0 \psi_s + h_d \psi_u^0
\psi_s) - (\k s+\half \mu_s) \psi_s \psi_s \nn \\
&& + \frac{i g_1}{\sqrt{2}}\, \l_1 (h_u \psi_u^0 - h_d \psi_d^0)
   - \frac{i g_2}{\sqrt{2}}\, \l_2^3 (h_u \psi_u^0 - h_d \psi_d^0) ,
\eea
where $\l_1$ is the $U(1)_Y$ gaugino and $\l_2^3$ is the neutral $SU(2)$ gaugino.
In the basis $\psi^0 = (-i\l_1 , -i\l_2, \psi_u^0, \psi_d^0, \psi_s)$ we obtains
\beq
{\cal L} = - \half \psi^0 {\cal M}_0 (\psi^0)^T + \mathrm{h.c.} ,
\eeq
where
\beq
{\cal M}_0 =
\left( \ba{ccccc}
M_1 & 0 & \frac{g_1 h_u}{\sqrt{2}} & -\frac{g_1 h_d}{\sqrt{2}} & 0 \\
& M_2 & -\frac{g_2 h_u}{\sqrt{2}} & \frac{g_2 h_d}{\sqrt{2}} & 0 \\
& & 0 & -(\mu+\l s) & -\l h_d \\
& & & 0 & -\l h_u \\
& & & & 2 \k s+\mu_s
\ea \right) . \label{neutralino_matrix}
\eeq
Diagonalizing this mass matrix, one obtains 5 mass eigenstates (ordered in mass)
\beq
\tilde\chi^0_i = N_{ij} \psi^0_j .\label{singlino}
\eeq
\end{itemize}

\section{ Explanation of Pamela and implication on Higgs decays}
\label{sec3}
\begin{figure}[htbp]
\epsfig{file=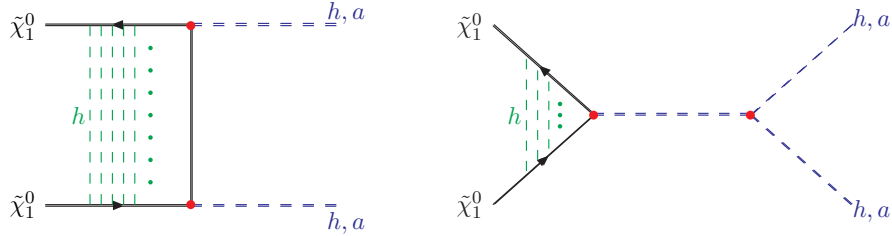,width=12cm} \vspace*{-0.5cm} \caption{Feynman
diagrams for singlino dark matter annihilation where Sommerfeld
enhancement is induced by exchanging $h$.} \label{fig1}
\end{figure}

In our study the lightest CP-odd neutral Higgs boson $a_1$ is singlet-dominant,
while for the CP-even neutral Higgs bosons the lightest one $h_1$ is singlet-dominant
and the next-to-lightest $h_2$ is doublet-dominant. We use the notation:
\beq
a\equiv a_1, ~~~~h\equiv h_1, ~~~~h_{SM}\equiv h_2. 
\label{definition}
\eeq
As discussed in \cite{Hooper:2009gm}, when the lightest neutralino $\tilde\chi^0_1$
in Eq.(\ref{singlino}) is singlino-dominant, it can be a perfect candidate for
the dark matter. As shown in Fig.\ref{fig1},
such singlino dark matter annihilates to a pair of light
singlet Higgs bosons followed by the decay $h\to aa$
($h$ has very small mixing with the Higgs doublets
and thus has very small couplings to the fermions).
In order to decay dominantly into muons,
$a$ must be light enough.
Further, in order to induce the Sommerfeld enhancement, $h$ must also be
light enough.
From the superpotential term $\kappa \hat S^3$ we know
that the couplings $h\tilde\chi^0_1\tilde\chi^0_1$ and $a\tilde\chi^0_1\tilde\chi^0_1$
are proportional to $\kappa$. To obtain the relic density of the dark matter, $\kappa$
should be ${\cal O}(1)$.

Since $h, a$ must be singlet-dominant and $\tilde\chi^0_1$ must be
singlino-dominant, this implies small mixing between singlet and
doublet Higgs fields. From the superpotential in
Eq.(\ref{superpotential}) we see that this means the mixing
parameter $\l$ must be small enough. On the other hand, the
smallness of $\l$ is also required by the lightness of  $h_1$ and
$a_1$ whose masses are approximately given by \bea {\cal M}_{h,33}
& \simeq & \k s\left[\l (A_\l+\mu_s) \frac{h_u h_d}{\k s^2}
-\l\frac{\mu}{\k s^2}(h_u^2+h_d^2)
+ \left(A_\k + 4 \k s+3\mu_s-\frac{C\eta}{\k s^2}\right)\right], \label{h1_mass}\\
{\cal M}_{a,22} & \simeq & \k s\left[\l (A_\l+\mu_s) \frac{h_u h_d}{\k s^2}
-\l\frac{\mu}{\k s^2}(h_u^2+h_d^2)
-\left(3A_\k+\mu_s+\frac{C\eta}{\k s^2}+\frac{2B_s\mu_s}{\k s}\right)
 \right] .\label{a1_mass}
\eea

In the following we scan over the parameter space. We modify the
package NMSSMTools \cite{nmssmtools} and use it in our
calculations. As discussed above, $\l$ must be small enough in
order to get a singlino-dominant $\tilde \chi^0_1$ and
singlet-dominant $h,a$ (we checked from our scan that $\l$ must be
smaller than 0.01 in order to get $m_{a}<0.5$ GeV and $m_{h}<20$
GeV). So in our following scan we fix $\l=10^{-3}$. Further, $\k$
is taken as $0.5$, and for the squark sector the soft masses and
the trilinear terms are fixed as 500 GeV. Other parameters vary in
the ranges: \bea
&& -500{\rm ~GeV}<C,~\mu,~\mu_s,~B,~A_\l,~M_1,~M_2<500{\rm ~GeV}\nn\\
&& -(500{\rm ~GeV})^2<\eta<(500{\rm ~GeV})^2,~~s<500{\rm ~GeV},
~~ 2<\tan\beta<40 .
 \label{scan}
\eea
In order to get small ${\cal M}_{h,33}$ and  ${\cal M}_{a,22}$,
the third terms in Eqs.(\ref{h1_mass},\ref{a1_mass}), which are not
suppressed by a small $\l$, must also be small.
Therefore, in our scan we require parameters $A_\k$ and $B_s$ to be
in the ranges:
\bea
A_\k  &\in& \left(-4 \k s-3\mu_s+\frac{C\eta}{\k s^2}\right) \pm 20{\rm GeV},\\
2B_s \mu_s  &\in& \left(-3A_\k \k s -\mu_s\k s-\frac{C\eta}{s}\right)
    \pm (3{\rm GeV})^2
\eea
In addition, we consider the following constraints:
\begin{itemize}
\item[(i)]  The constraints from the LEP experiments, which include the LEP1 bound on
            invisible $Z$ decay and the LEP2 direct searches for Higgs bosons;
\item[(ii)] $m_{a_1}<0.5$ GeV;
\item[(iii)] The singlino-like $\tilde\chi^0_1$ to give the dark matter relic density
             $\Omega_{\tilde\chi^0_1}h^2$ in the range 0.01-0.2, which can be calculated
            from the approximate formula \cite{Hooper:2009gm}
\beq
\Omega_{\tilde\chi^0_1}h^2\sim 0.1\times\left(\frac{0.5}{\k}\right)^2
\left(\frac{m_{\chi^0}}{200{\rm GeV}}\right)^2 .
\eeq
\end{itemize}
To calculate the Sommerfeld enhancement we follow \cite{sommerfeld2} to
numerically solve the Schr\"odinger equation
\begin{equation}
  -\frac{1}{2M}\frac{d^2}{dr^2}\chi+V(r)\chi=\frac{k^2}{2M}\chi \label{sde3}
\end{equation}
 with the boundary condition ($r\to\infty$)
\begin{equation}
  \chi(r)\to \sin(kr+\delta) ,
\end{equation}
where $M$ and $k$ are respectively the mass and momentum of the
dark matter particle. $V(r)$ is the Yukawa potential induced by
exchanging $h$ and is given by
\begin{equation}
  V(r)=-\frac{\kappa^2}{2\pi}\frac{e^{-m_{h} r}}{r}.
\end{equation}
The Sommerfeld enhancement is then given by
\begin{equation}
 T=\left|\frac{\frac{d\chi}{dr}(0)}{k}\right|^2.
\end{equation}
\begin{figure}[htbp]
\epsfig{file=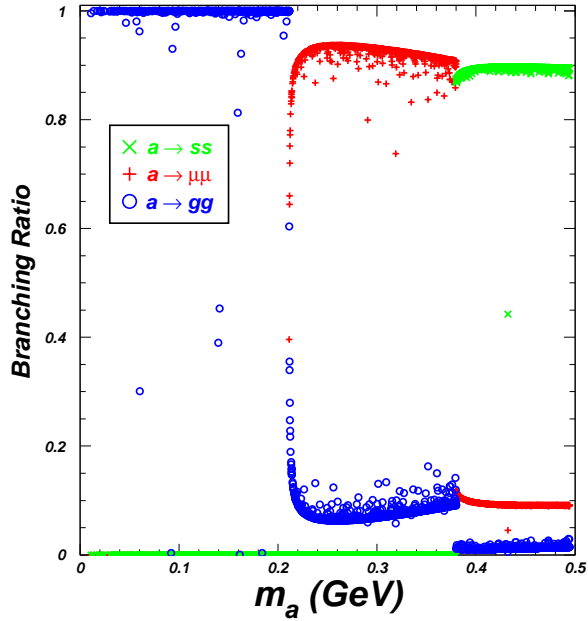,width=8cm}
\vspace*{-0.7cm}
\caption{The scatter plots showing the decay branching ratios
$a\to\mu^+\mu^-$ (muon), $a\to gg$ (gluon)
and $a\to s\bar s$ ($s$-quark) versus $m_a$
         for $\l=10^{-3}$.}
\label{fig2}
\end{figure}
\begin{figure}[htbp]
\epsfig{file=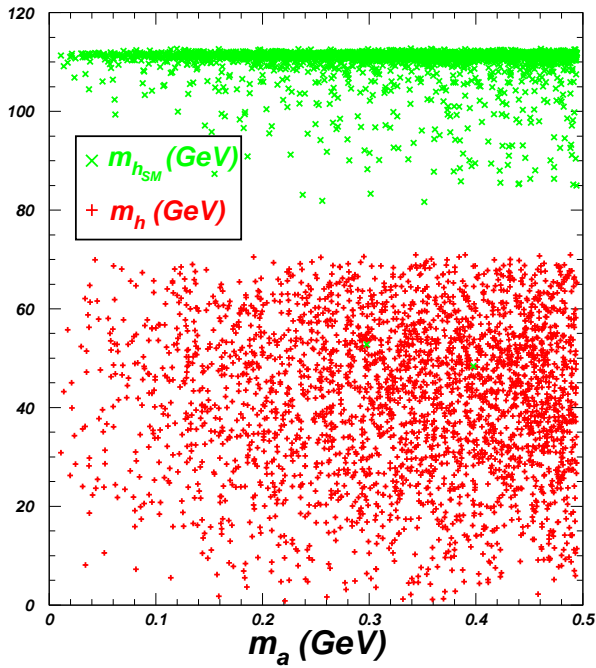,width=8cm}
\vspace*{-0.7cm}
\caption{Same as Fig.\ref{fig2}, but showing  $m_h$ and $m_{h_{SM}}$ versus $m_a$.}
\label{fig3}
\end{figure}
\begin{figure}[htbp]
\epsfig{file=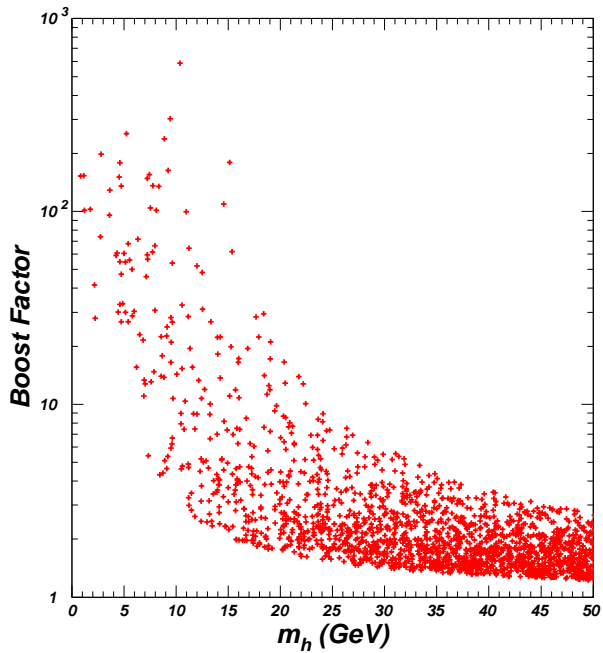,width=8cm} \vspace*{-0.7cm} \caption{Same as
Fig.\ref{fig2}, but showing the Sommerfeld enhancement factor
 induced by $h$.}
\label{fig4}
\end{figure}
\begin{figure}[htbp]
\epsfig{file=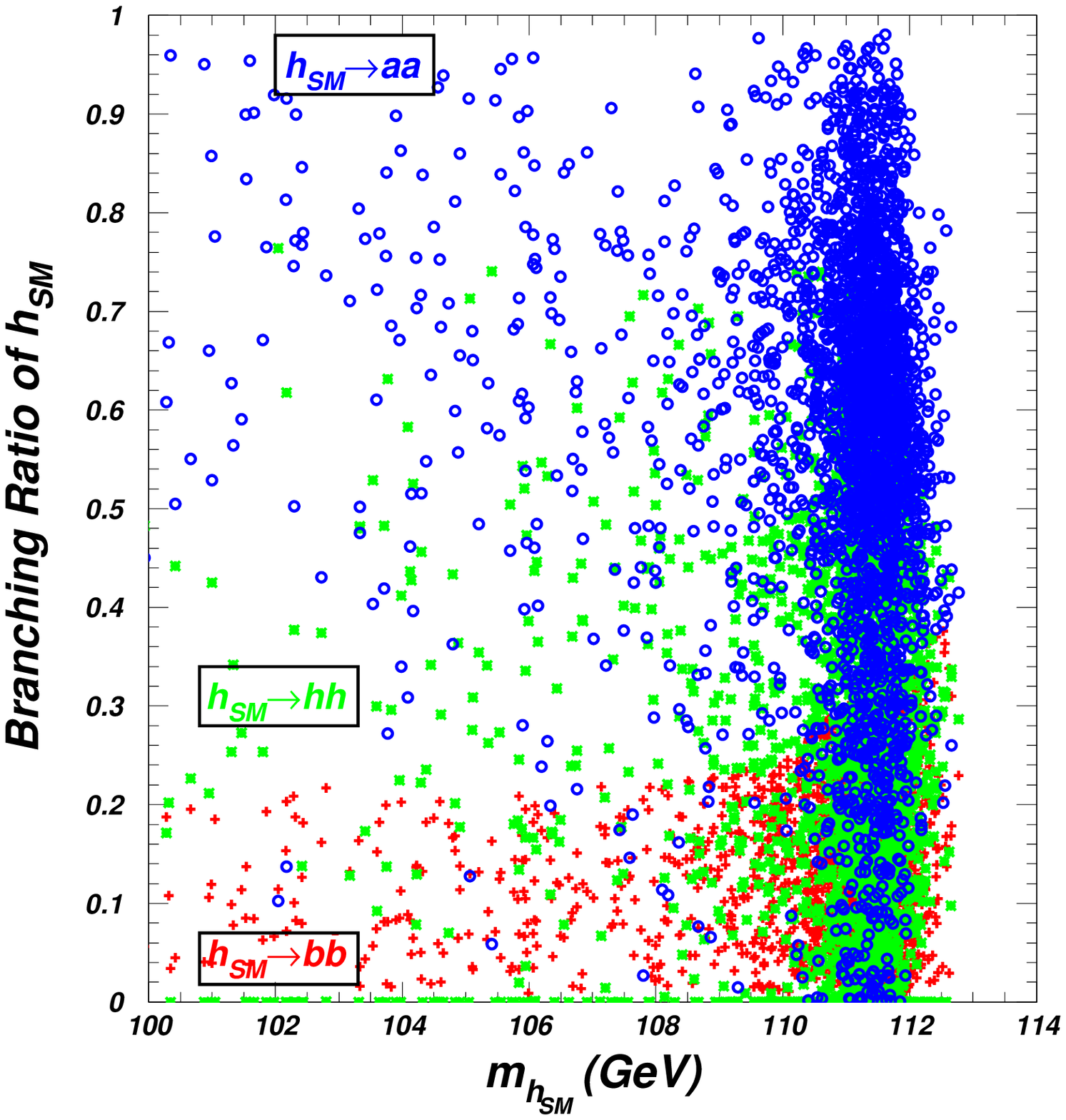,width=8cm}
\vspace*{-0.7cm}
\caption{Same as Fig.\ref{fig2}, but showing the branching ratio of $h_{SM}$ decays.
The '$\circ$' (blue), '$\times$' (green) and  '$+$' (red) denote the branching ratios
of $h_{SM}\to a a$, $h_{SM}\to h h$ and  $h_{SM}\to b \bar b$, respectively.}
\label{fig5}
\end{figure}
\begin{figure}[htbp]
\epsfig{file=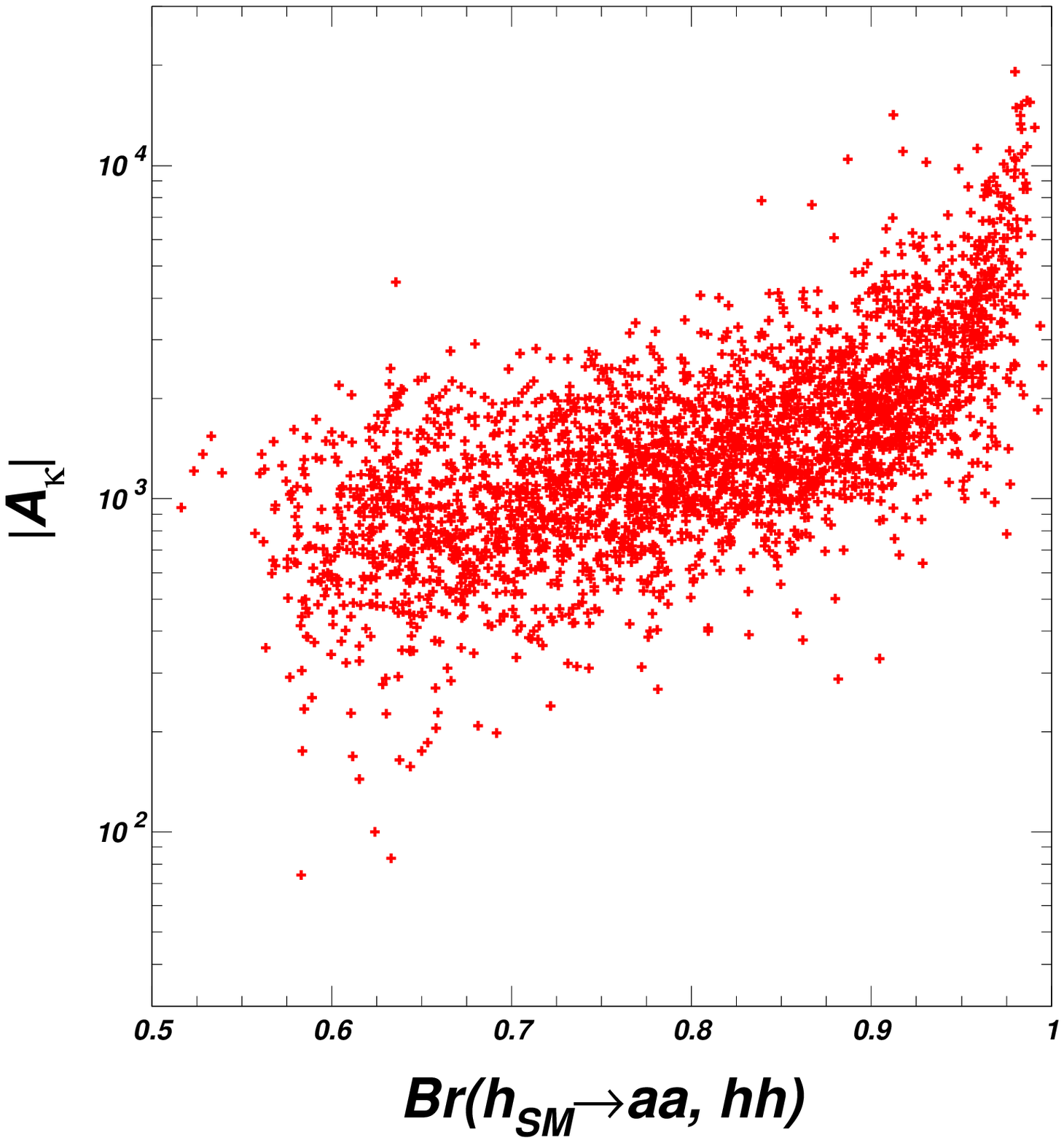,width=8cm}
\vspace*{-0.7cm}
\caption{Same as Fig.\ref{fig2}, but showing $|A_\k|$ versus the branching ratio
         of $h_{SM}\to a a, hh$.}
\label{fig6}
\end{figure}
The survived points are displayed in different planes in
Figs.\ref{fig2}-\ref{fig6}. We see from Fig.\ref{fig2} that in the
range $2m_{\mu}<m_{a}<2m_{\pi}$, $a$ decays dominantly into muons.
From Fig.\ref{fig3} it is clear  that  $h$ can be as light as a
few GeV, which is light enough to induce the necessary Sommerfeld
enhancement as shown in  Fig.\ref{fig4}. In the calculation of the
Sommerfeld enhancement, we assumed the dark matter move with a
velocity $150{\rm ~km/s}$.

The fit to Pamela result has been given in \cite{Hooper:2009gm}.
As shown in Table I in \cite{Hooper:2009gm}, for the parameter
space in Figs.\ref{fig2}-\ref{fig4} with $2m_{\mu}<m_{a}<2m_{\pi}$
and $m_h$ as light as a few GeV (so the Sommerfeld enhancement
factor is large enough), the Pamela positron excess can be
naturally explained.

In Fig.\ref{fig5} we show the branching ratios of $h_{SM}$ decays.
We see that in the allowed parameter space $h_{SM}$ tends to decay into $aa$
or $hh$ instead of $b\bar b$. This can be understood as following.
The MSSM parameter space
is stringently constrained by the LEP experiments if $h_{SM}$ is relatively light and
decays dominantly
to $b\bar b$, and to escape such stringent constraints $h_{SM}$ tends to have
exotic decays into  $aa$ or $hh$. As a result, the allowed parameter space tends
to favor a large $A_\kappa$, as shown in Fig.\ref{fig6}, which greatly enhances the 
couplings $h_{SM}aa$ and $h_{SM}hh$ through the soft term $\kappa A_\kappa S^3$ 
although $S$ has a small mixing with the doublet Higgs bosons. 
Such an enhancement can be easily seen.
Take the coupling  $h_{SM}hh$ as an example. 
The soft term $\kappa A_\kappa S^3$ gives a term $\kappa A_\kappa S^3_R$ which
then gives the interaction $\kappa A_\kappa ~U_{13}^2 U_{23}~ h_{SM}hh$
because $S_R=U_{13}h_1+U_{23}h_2+U_{33}h_3$ with $h_1\equiv h$ and 
$h_2\equiv h_{SM}$ (see Eqs.\ref{rotation} and \ref{definition}). 
Although the mixing $U_{13}^2U_{23}$ is small for 
a small $\lambda$, a large $A_\kappa$ can enhance the coupling  $h_{SM}hh$. 
 
The SM-like Higgs boson $h_{SM}$ will be intensively searched at the LHC
and its dominant decay mode in the MSSM is $b\bar b$. In the singlet extension
of the MSSM, its dominant decay mode may be changed to $a a$ or $h h$, as shown
in our above results. Such new decay modes will give a
multi-muon signal for $h_{SM}$ at the LHC, i.e.,
$h_{SM}\to aa\to 4 \mu$ or $h_{SM}\to hh\to 4 a \to 8 \mu$.
So the phenomenology of $h_{SM}$ will be quite different from the MSSM predictions.

Finally, we make some remarks regarding our results:
\begin{itemize}
\item[(1)] The recent D0 search for $h\to a a\to
4\mu {\rm ~or~ } 2\mu 2\tau$ channel obtained null results, which
constrained the parameter space for the CP-odd Higgs $a$ in
the mass range of 3.6-9.5 GeV \cite{D0-note}. 
But they do not constrain the parameter space considered 
in our analysis because we considered a much lighter CP-odd Higgs $a$ 
with a mass below 0.5 GeV. 
Also, as pointed in
\cite{Hooper:2009gm}, such a light $a$ is allowed by $\Upsilon (3s)
\rightarrow \gamma a \rightarrow \gamma \mu^+ \mu^-$
\cite{upsilon} and $K^+ \rightarrow \pi^+ a \rightarrow \pi^+
\mu^+ \mu^-$ \cite{k-decay} because in our scenario $a$ is over
dominated by singlet. 
\item[(2)] In the allowed parameter space
displayed in our results, the mass of the SM-like Higgs boson
$h_{SM}$ is rather below its theoretical upper bound (about 135
GeV in the MSSM). The reason is that, in order to push up its
mass, the loop effects of heavy stops are needed ( note that in
the singlet extension the tree-level upper bound can be enhanced
by a term proportional to $\l$, which is very small in our
scenario). In our calculations the soft mass parameters in the
squark sector are fixed to be 500 GeV and hence the stops are not
heavy enough to push the mass of  $h_{SM}$ up to 135 GeV. Of
course, we can choose heavy stops to push up the mass of
$h_{SM}$, in which case the allowed parameter space displayed in
our results (with a relatively light $h_{SM}$ decaying dominantly
into $aa$ or $hh$) can still survive. 
\item[(3)] For the specified
singlet extensions like nMSSM and NMSSM \cite{nmssm}, the
explanation of Pamela and relic density through Sommerfeld
enhancement is not possible. The reason is that the parameter
space of such models is stringently constrained by various
experiments and dark matter relic density \cite{limit-nmssm}, and,
as a result, the neutralino dark matter may explain either the
relic density or Pamela, but impossible to explain both via
Sommerfeld enhancement \cite{dm-nmssm}. For example, in the nMSSM
various experiments and dark matter relic density constrain the
neutralino dark matter particle in a narrow mass range
\cite{limit-nmssm}, which is too light to explain Pamela.
\end{itemize}

\section{Summary}\label{sec4}
The singlino-like dark matter in the MSSM extended by a singlet
Higgs superfield can give a perfect explanation for both  the
relic density and  the Pamela result through the
Sommerfeld-enhanced annihilation into singlet Higgs bosons ($a$ or
$h$ followed by $h\to a a$) with $a$ being light enough to decay
dominantly to muons. In this work we analyzed the parameter space
allowed by such a dark matter explanation and also considered the
constraints from the LEP experiments. We found that although the
light singlet Higgs bosons have small mixings with the Higgs
doublets in the allowed parameter space, their couplings with the
SM-like Higgs boson $h_{SM}$ can be enhanced by the soft parameter
$A_\kappa$ and, in order to meet the stringent LEP constraints,
the $h_{SM}$ tends to decay into the singlet Higgs pairs $aa$ or
$hh$ instead of $b\bar b$, which will give a multi-muon signal for
$h_{SM}$ produced at the LHC, $h_{SM}\to aa\to 4 \mu$ or
$h_{SM}\to hh\to 4 a \to 8 \mu$.

\section*{Acknowledgment}
We thank Xiaojun Bi, Junjie Cao, Tao Han and Chun Liu for discussions.
This work was supported in part by the National Natural
Science Foundation of China under grant Nos. 10821504, 10725526 and 10635030,
and by the Natural Science Foundation of Beijing under grant No. 1072001.

\end{document}